\documentstyle[sprocl]{article}
\def\NPB{{\em Nucl. Phys.} B }
\def\PLB{{\em Phys. Lett.} B }
\def\PRL{{\em Phys. Rev. Lett.} } 
\def\PRD{{\em Phys. Rev.} D }
\def\PR{{\em Phys. Rev.} } 
\def\DKD{{\em Det. Kong. Danske Videnskabernes Selskab
Mat.-fysiske Meddelelser} } 
\def\PRS{{\em Proc. Roy. Soc. (London)} A }
\def\JETP{{\em JETP Lett.} }

\def\ga{\gamma}
\def\de{\delta}
\def\ep{\epsilon}

\def\la{\lambda}

\def\ph{\phi}

\def\ch{\chi}
\def\ps{\psi}

\def\Ga{\Gamma}
\def\De{\Delta}

\def\La{\Lambda}

\def\cl{{\cal L}}
\def\fr#1#2{{{#1} \over {#2}}}
\def\prt{\partial}

\def\vev#1{\langle {#1}\rangle}

\def\frac#1#2{{\textstyle{{#1}\over {#2}}}}

\def\lsim{\mathrel{\rlap{\lower4pt\hbox{\hskip1pt$\sim$}}
    \raise1pt\hbox{$<$}}}
\def\gsim{\mathrel{\rlap{\lower4pt\hbox{\hskip1pt$\sim$}}
    \raise1pt\hbox{$>$}}}
\def\sqr#1#2{{\vcenter{\vbox{\hrule height.#2pt
         \hbox{\vrule width.#2pt height#1pt \kern#1pt
         \vrule width.#2pt}
         \hrule height.#2pt}}}}

\def\Re{\hbox{Re}\,}
\def\Im{\hbox{Im}\,}

\newcommand{\beq}{\begin{equation}}
\newcommand{\eeq}{\end{equation}}
\newcommand{\bea}{\begin{eqnarray}}
\newcommand{\eea}{\end{eqnarray}}
\newcommand{\rf}[1]{(\ref{#1})}

\begin{document}

\begin{flushright}
IUHET 360\\
March 1997
\end{flushright}
\bigskip

\title{TESTING CPT SYMMETRY}
\author{V.A. KOSTELECK\'Y}
\address{Physics Department, Indiana University,\\
Bloomington, IN 47405, U.S.A.}

\maketitle\abstracts{
In this talk,
I review the possibility that CPT and Lorentz symmetry
might be spontaneously broken in nature by effects
originating in a theory beyond the standard model,
and I discuss some existing and future experimental tests.}

\section{Introduction}

This talk provides a short review of some theoretical
and experimental results 
relevant to the possibility that observable CPT violation
could be generated in a theory beyond the standard model.

The product CPT of the discrete transformations involving 
charge conjugation, parity reflection, and time reversal
is predicted theoretically to be an exact invariance of 
local relativistic field theories of point 
particles.\cite{cpt1}$^{\!-\,}$\cite{cpt7}
This prediction agrees with many experimental tests
performed in various systems,
some to considerable accuracy.\cite{pdg}
The generality of the theoretical prediction
and the existence of high-precision experimental tests
means that CPT violation is an interesting possible signature 
for new physics that could emerge from a fundamental theory
such as strings.\cite{kp1}$^{\!-\,}$\cite{kp3}

Among existing approaches to a fundamental theory,
strings currently remain the most promising avenue
for the development of a complete and consistent 
quantum description 
of all fundamental interactions and particles.
For string theory,
standard assumptions in the proofs of the CPT theorem 
are open to question because strings are extended objects.
In Sec.\ 2, I outline a mechanism for 
spontaneous 
Lorentz\cite{ks}
and 
CPT\cite{kp1,kp2}
violation that arises in this context
and mention some theoretical tests 
of these ideas.\cite{ks2,kp4}

If indeed spontaneous CPT violation occurs in nature,
then it can be studied at presently accessible energies
through an effective theory that allows for  
suitable CPT-violating interactions.
Section 3 outlines some results leading
to a general extension of the standard model
incorporating additional interactions
that could arise from spontaneous 
Lorentz and CPT violation.\cite{kp3,cksm}
These terms maintain the known gauge symmetries
and are power-counting renormalizable.

The standard-model extension can be used 
to investigate consequences of spontaneous
CPT and Lorentz breaking.
For example,
experiments that measure or bound
the coefficients of the additional terms in the
standard-model extension
provide quantitative
constraints on possible CPT and Lorentz violation in nature.
Section 4 outlines some results involving CPT tests using 
neutral-meson oscillations.
The neutral-meson systems are particularly interesting
because their interferometric nature
makes them exceptionally sensitive to CPT violation.
Tests of spontaneous CPT violation are possible
in the $K$ system,\cite{kp1,kp2,kp3}
the two $B$ systems,\cite{kp3,ck1,kv}
and the $D$ system.\cite{kp3,ck2}

Tests of CPT are also feasible in other systems.
Section 5 summarizes some results 
concerning CPT studies using measurements of 
the electron and positron anomalous magnetic moments\cite{bkr}
and the possibility that
CPT violation might play a significant role 
in baryogenesis.\cite{bckp}

The minuscule spontaneous CPT and Lorentz breakings
that are considered in this talk 
can be understood completely within 
conventional quantum mechanics.
Violations of conventional quantum mechanics 
that might lead to CPT breaking
have been proposed as possibly 
arising in the context of 
quantum gravity.\cite{qg1}$^{\!-\,}$\cite{qg3}
Experiments in the $K$ system would produce 
very different signatures
for the two kinds of CPT breaking.\cite{qg4,qg5}

\section{Spontaneous Lorentz and CPT Violation}

Suppose the fundamental theory of nature
is dynamically Lorentz invariant
but involves more than four spacetime dimensions.
Since we observe only four dimensions, 
it is plausible that the higher-dimensional Lorentz group
is spontaneously broken.

In string theory,
which is naturally formulated in higher dimensions,
a mechanism exists that could generate spontaneous
Lorentz violation.\cite{ks}
There exist interactions in string field theory,
emerging from string nonlocality,
that are compatible with string gauge invariances
and the infinite number of particle fields.
Comparable interactions are absent in
conventional four-dimensional renormalizable gauge theories.
These stringy interactions can destabilize the 
static potentials for Lorentz tensor fields
if certain scalars develop expectation values.
The stable solution of the theory
may then involve nonzero expectation values
for some Lorentz tensor fields,
which means Lorentz invariance is spontaneously broken.
In particular,
any expectation values involving
tensors with an odd number of spacetime indices
spontaneously break CPT.\cite{kp1,kp2}

The above ideas can be examined directly
using the string field theory of the open bosonic string.
The set of extrema of the action can systematically  
be established in a level-truncation scheme.\cite{ks2,kp4}
By allowing only particle fields below a specified 
level number $N$,
the action and then the equations of motion
can be analytically derived. 
Among the solutions to the equations of motion 
are ones that break Lorentz and CPT invariance.
These can be determined  
and compared with analogous ones 
obtained for different values of $N$.
The existence of a corresponding extremum 
of the full theory is plausible
if the solutions not only persist 
but also appear to converge to a definite set 
of nonzero expectation values
as $N$ increases.
Under certain circumstances,
it has been possible to use symbolic manipulation routines
to examine more than 20,000 nonvanishing terms
in the static potential obtained from the action.
The resulting solutions exhibit
Lorentz and CPT properties 
that are compatible with those expected  
from more general considerations
of the theoretical mechanism.

\section{Standard-Model Extension}

It is natural to ask if the above mechanism 
for spontaneous Lorentz and CPT violation
generates observable effects at low energies.
This would seem plausible theoretically,
as there is no evident reason why
four dimensions would be preferred.
Since neither Lorentz nor CPT violations
have been observed,
any effects at the level of the standard model 
must be highly suppressed.

{}From the perspective of a fundamental theory,
the standard model can be regarded as an
effective low-energy model
at the electroweak scale $m_{\rm ew}$.
If the scale controlling the fundamental theory
is the Planck mass $m_{\rm Pl}$,
then there is a natural suppression factor
$r \sim m_{\rm ew}/m_{\rm Pl} \simeq 10^{-17}$ 
for Planck-scale effects at the electroweak scale. 
Note that with a suppression at this level,
relatively few Lorentz- and CPT-violating effects 
are potentially observable.
Some possible signals are discussed in later sections.

The spontaneous CPT and Lorentz violation might 
thus produce suppressed low-energy contributions 
to the standard model.
For example,
a generic contribution to the fermionic sector 
of the standard model could arise 
from terms in a compactified string theory
of the form\cite{kp2,kp3}
\beq
\cl \sim \fr {\la} {M^k} 
\vev{T}\cdot\overline{\ps}\Ga(i\prt )^k\ch
+ {\textstyle h.c.}
\quad .
\label{a}
\eeq
Terms of this type are Lorentz and CPT breaking by virtue 
of nonzero expectation values of Lorentz tensors $T$.
The tensors are coupled to
four-dimensional fermions $\ps$ and $\ch$ 
via derivatives $i\prt$
and a gamma-matrix structure $\Ga$.
If $\la$ is taken as a dimensionless coupling constant,
then a suitable power of one or more large mass scales $M$ 
(possibly the Planck or compatification scales)
must also appear.

It is of interest to determine a general extension
of the standard model that allows for all types of effects 
that could in principle arise 
from spontaneous Lorentz and CPT breaking.
These include, 
for example,
extra terms of the form \rf{a}
when the fermions $\ps$ and $\ch$ 
are identified with appropriate
fermions in the standard model.
Imposing the usual 
SU(3) $\times$ SU(2) $\times$ U(1)
gauge invariance 
and requiring power-counting renormalizability
significantly restricts the possibilities.
The general extension of the standard model,
including Lorentz-breaking terms both with and
without CPT breaking,
has been established.\cite{cksm}
The derivation includes a framework
for treating theoretically the effects
of CPT and Lorentz breaking.
It appears that several of the usual difficulties 
are avoided by the spontaneous nature of the breaking,
which reflects the noninvariance of solutions
to the equations of motion rather than 
dynamical violations in the action.

The remainder of the talk
summarizes some possible observable consequences
of the standard-model extension.
To date, 
these have been most fully explored in
the neutral-meson systems,
as is outlined in the next section.
Other possible signals exist,
however,
as is briefly reviewed in Sec.\ 5.

\section{Tests in Neutral-Meson Systems}

The neutral-meson systems provide excellent
hunting grounds for CPT violation
because their interferometric nature
makes them potentially sensitive to Planck-scale effects. 
There are four neutral-meson systems to consider:\cite{kp3}
$K$, $D$, $B_d$, and $B_s$.
In what follows,
the symbol $P$ is used to denote any one of these.
In the next subsection
some theoretical considerations are given,
while the subsequent one summarizes
the current experimental situation.

\subsection{Theory}

Among the terms in the extension of the standard model 
are ones involving quark fields.
For example,
if $\ps$ and $\ch$ are regarded as valence quarks in a 
neutral meson $P$,
then the terms \rf{a} modify 
the $2\times 2$ effective hamiltonian $\La$ governing the 
$P$-meson time evolution.

{}From a purely phenomenological viewpoint,
independent of any underlying theory
except the assumptions of conventional quantum mechanics,
two types of (indirect) CP violation
might occur in $\La$.
One is given by the usual CP violating parameter $\ep_P$,
which violates T but preserves CPT.
The other is given by a complex CP violating parameter $\de_P$,
which preserves T but breaks CPT.

Within the usual standard model,
nonzero parameters $\ep_P$ for different $P$ can be
understood in terms of the CKM matrix.
Similarly,
the CPT-violating extension of the 
standard model mentioned above
permits an understanding of 
possible nonzero values of the parameters $\de_P$
for different $P$.

Within the framework of spontaneous CPT and Lorentz breaking,
the parameter $\de_P$ for a given $P$ system
is given by\cite{kp2,kp3}
\beq
\de_P = i 
\fr{h_{q_1} - h_{q_2}}
{\sqrt{\De m^2 + \De \ga ^2/4}}
e^{i\hat\ph}
\quad ,
\label{b}
\eeq
where 
the experimental observables
$\De m$ and $\De\ga$ are mass and rate differences,
while $\hat\ph = \tan^{-1}(2\De m/\De \ga)$.
The parameters $h_{q_j}=r_{q_j}\la_{q_j}\vev{T}$
are controlled by the fundamental theory
and by effects $r_{q_j}$ of the quark-gluon sea.
They originate from terms in the standard-model extension
of the form \rf{a}.

The underlying fundamental theory 
and the standard-model extension are 
assumed to be hermitian.
It follows that the $h_{q_j}$ are real,
which determines a relation
involving experimental observables
between the real and imaginary
components of $\de_P$:
\beq
\Im \de_P = \pm \fr{\De\ga}{2\De m} ~\Re\de_P
\quad . 
\label{c}
\eeq
If indeed the suppression ratio for Planck-scale effects
is $r\simeq 10^{-17}$,
then detection of direct CPT violation
in the $P$-meson decay amplitudes is excluded.
The condition \rf{c}
is therefore expected to be a key signature
for CPT violation within the present framework.

In the context of the standard-model extension
the CPT-violating couplings could be very different
for the different quarks,
in analogy to the Yukawa couplings 
that vary over some six orders of magnitude.
Equivalently,
the dimensionless coupling constants $\la_{q_j}$
of \rf{a} could change with quark flavor $q_j$.
It is therefore possible that 
the CPT-violating quantities $\de_P$ 
determined in \rf{b}
could vary significantly for different $P$ mesons.

Since $\de_P$ might differ for distinct $P$,
it is important to test CPT experimentally in
more than one neutral-meson system.
There are also other implications.
For example,
the relatively weak bounds 
presently available on $B_d$-meson CP violation
still allow the possibility
that conventional CP violation through $\ep_{B_d}$
is \it smaller \rm than
CP violation through the CPT-violating parameter $\de_{B_d}$.
This would induce interesting experimental signals
in the proposed $B$ factories.

\subsection{Experiment}

Indirect CP violation in a given $P$ system,
including both indirect T and CPT violations,
can be experimentally studied with
correlated $P$-$\overline P$ pairs arising
from quarkonium decays
or with uncorrelated tagged $P$ mesons.
The relevant experimental variables
are appropriate asymmetries of decay probabilities
into different final states.
Suitable asymmetries,
including ones with time dependences,
now exist for all $P$
for both correlated and uncorrelated situations. 
They can be adopted for detailed Monte-Carlo simulations
of realistic experimental data
including acceptances and background effects,
as well as for relatively straightforward analytical estimates
of CP reach.
What follows is a brief summary of the status of
indirect CPT tests for each possible $P$ system.
More details can be found in the 
literature.\cite{kp1}$^{\!-\,}$\cite{ck2}

The best neutral-meson bound on CPT violation 
has been established in the $K$ system. 
Published limits\cite{pdg,expt1,expt2}
on $|\de_K|$ are of order $10^{-3}$.
Improved bounds are expected in the near future,
coming from completed experiments 
(e.g., CPLEAR\cite{expt3} at CERN) 
or ongoing ones
(e.g., KTeV at Fermilab).
There are also good future prospects 
from $\ph$ factories,\cite{expt4}
designed to produce a large flux 
of correlated $K$-$\overline{K}$ pairs.

Unlike the $K$ system,
mixing has not been experimentally seen in the $D$ system.
Moreover,
theoretical estimates are uncertain,
possibly by orders of magnitude,
due to the combination of strong dispersive effects
and the possibility of contributions from 
beyond the standard model.
Under favorable conditions
some tests of CPT symmetry may be feasible 
in the $D$ system using attainable experimental methods
and possibly even with current data.
The increased statistics available from
future facilities\cite{expt5}
offers interesting possibilities for developing CPT bounds.

Since the $B_d$ system involves the $b$ quark,
it has the potential to generate
the largest CPT violation
and is therefore particularly interesting for CPT tests.
Until very recently,
no limit on $\de_{B_d}$ had been established,
although enough data for this purpose have 
already been obtained in the CERN LEP experiments 
and in CLEO experiments at 
Cornell.\cite{kv}
The OPAL collaboration at CERN has 
now placed a limit\cite{expt6}
on $\Im\de_{B_d}$ of about $2 \times 10^{-2}$.
The CLEO data could be used to bound 
$\Re\de_{B_d}$ at the level of about 10\%. 
The $B$ factories and other $B$-dedicated experiments
currently under development 
are expected to improve this bound considerably.

\section{Other Effects}

The discussion of spontaneous Lorentz and CPT violation
in the context of a fundamental theory 
and the corresponding standard-model extension 
suggest the possibility of signals of Lorentz and CPT
violation in systems other than neutral mesons.
In this section,
a short outline is
given of a few of these that have been elucidated.

If conditions are suitable,
terms of the form \rf{a}
might provide an acceptable mechanism for
baryogenesis in thermal equilibrium.\cite{bckp}
The mechanism could produce a large asymmetry
at grand-unification scales
that is subsequently reduced to the experimental value,
for example,
through sphaleron dilution.
In contrast,
conventional approaches require nonequilibrium processes
and C- and CP-breaking interactions,\cite{ads}
which in grand-unified theories are chosen to match
the observed baryon asymmetry 
without relation to experimentally measured CP violation
within the standard model.

Other possible observable signals could arise in 
the context of quantum electrodynamics.\cite{cksm,bkr}
For example,
experiments determining the difference between
the anomalous magnetic moments of the electron and positron
have the potential to constrain tightly CPT violation.
Indeed,
the conventional figure of merit adopted in these 
experiments provides a misleading measure of CPT
violation.\cite{bkr}
A more appropriate figure of merit
suggests that bounds on CPT violation 
could be placed on leptonic systems
comparable to those in neutral mesons.
Other bounds may be imposed
from known properties of photons.\cite{cksm}
A variety of experimental tests is essential
because the effects in the different sectors 
are controlled by
different parameters in the standard-model extension.

\section*{Acknowledgments} 

I thank Orfeu Bertolami, Robert Bluhm, Don Colladay, 
Rob Potting, Neil Russell, Stuart Samuel, 
and Rick Van Kooten for collaborations.
This work was supported in part
by the United States Department of Energy 
under grant number DE-FG02-91ER40661.


\section*{References}

\begin{thebibliography}{xx}

\bibitem{cpt1}
J. Schwinger, 
\PR 
{\bf 82} (1951) 914.

\bibitem{cpt2}
G. L\"uders, 
\DKD
{\bf 28}, no.\ 5 (1954).

\bibitem{cpt3}
J.S. Bell,
Ph.D.\ thesis (Birmingham University, England, 1954);
\PRS
{\bf 231} (1955) 479.

\bibitem{cpt4}
W. Pauli, in 
\it Niels Bohr and the Development of Physics, \rm
ed.\ W. Pauli,
(McGraw-Hill, New York, 1955), p.\ 30.

\bibitem{cpt5}
G. L\"uders and B. Zumino,
\PR 
{\bf 106} (1957) 385.

\bibitem{cpt6}
R.F. Streater and A.S. Wightman,
\it PCT, Spin and Statistics, and All That \rm
(Benjamin Cummings, Reading, 1964).

\bibitem{cpt7}
R. Jost,
{\it The General Theory of Quantized Fields}
(AMS, Providence, 1965).

\bibitem{pdg}
See, for example, R.M. Barnett {\it et al.},
{\it Review of Particle Properties,}
\PRD 
{\bf 54} (1996) 1.

\bibitem{kp1}
V.A. Kosteleck\'y and R. Potting,
\NPB 
{\bf 359} (1991) 545.

\bibitem{kp2}
V.A. Kosteleck\'y, R. Potting, and S. Samuel,
in 
\it Proceedings of the 1991 Joint International Lepton-Photon
Symposium and Europhysics Conference on High Energy Physics, \rm
eds.\ S. Hegarty et al.\
(World Scientific, Singapore, 1992); 
V.A. Kosteleck\'y and R. Potting,
{\it Gamma Ray--Neutrino Cosmology and Planck Scale Physics,} \rm
ed.\ D.B. Cline
(World Scientific, Singapore, 1993)
(hep-th/9211116).

\bibitem{kp3}
V.A. Kosteleck\'y and R. Potting,
\PRD 
{\bf 51} (1995) 3923.

\bibitem{ks}
V.A. Kosteleck\'y and S. Samuel,
\PRD 
{\bf 39} (1989) 683;
\it ibid., \rm
{\bf 40} (1989) 1886;
\PRL {\bf 63} (1989) 224;
\it ibid., \rm
{\bf 66} (1991) 1811.

\bibitem{ks2}
V.A. Kosteleck\'y and S. Samuel,
\NPB 
{\bf 336} (1990) 263;
\PRL 
{\bf 64} (1990) 2238;
\PRD 
{\bf 42} (1990) 1289.

\bibitem{kp4}
V.A. Kosteleck\'y and R. Potting,
\PLB 
{\bf 381} (1996) 389.

\bibitem{cksm}
D. Colladay and V.A. Kosteleck\'y,
\PRD 
{\bf 55} (1997) 6760;
Indiana University preprint IUHET 359 (1997).

\bibitem{ck1}
D. Colladay and V.A. Kosteleck\'y,
\PLB 
{\bf 344} (1995) 259.
 
\bibitem{kv}
V.A. Kosteleck\'y and R. Van Kooten,
\PRD 
{\bf 54} (1996) 5585.

\bibitem{ck2}
D. Colladay and V.A. Kosteleck\'y,
\PRD 
{\bf 52} (1995) 6224.

\bibitem{bkr}
R. Bluhm, V.A. Kosteleck\'y, and N. Russell,
\PRL 
{\bf 79} (1997) 1432.

\bibitem{bckp}
O. Bertolami, D. Colladay,
V.A. Kosteleck\'y, and R. Potting,
\PLB 
{\bf 395} (1997) 178.

\bibitem{qg1}
S.W. Hawking,
\PRD 
{\bf 14} (1976) 2460.

\bibitem{qg2}
D. Page,
\PRL 
{\bf 44} (1980) 301.

\bibitem{qg3}
R.M. Wald,
\PRD 
{\bf 21} (1980) 2742.

\bibitem{qg4}
J. Ellis, J.L. Lopez, N.E. Mavromatos, and D.V. Nanopoulos,
\PRD 
{\bf 53} (1996) 3846;
D.V. Nanopoulos, these proceedings.

\bibitem{qg5}
P. Huet, these proceedings.

\bibitem{expt1}
L.K. Gibbons 
{\it et al.}, 
\PRD 
{\bf 55} (1997) 6625.

\bibitem{expt2}
R. Carosi 
{\it et al.}, 
\PLB 
{\bf 237} (1990) 303.

\bibitem{expt3}
P. Kokkas, these proceedings.

\bibitem{expt4}
P. Franzini, these proceedings.

\bibitem{expt5}
D. Kaplan, these proceedings.

\bibitem{expt6}
OPAL Collaboration, R. Ackerstaff 
{\it et al.}, 
CERN preprint
CERN-PPE/97-036 (April 1997).

\bibitem{ads}
A.D. Sakharov,
\JETP 
{\bf 5} (1967) 24.

\end{thebibliography}
\end{document}